\documentclass[a4paper,11pt]{article}
\usepackage{aaskaiid}
\usepackage{xspace}
\usepackage{gensymb}
\usepackage{CJK}
\usepackage{orcidlink}
\newcommand{\oiline}{$\mathrm{[OI]}\,6300\,$\AA\xspace}
\newcommand{\Neline}{[NeII]\,12.81\,{\textmu m}\xspace}
\newcommand{\htline}{o-H$_2$\,2.12\,{\textmu m}\xspace}
\newcommand{\holine}{H$\mathrm{\alpha}$\xspace}
\newcommand{\prodimo}{\textsc{ProDiMo}\xspace}
\newcommand{\mocassin}{\textsc{mocassin}\xspace}
\newcommand\farcs{\mbox{$.\!\!^{\prime\prime}$}}

\title{Ionized gas emission in protoplanetary disks with the SKAO}
\ShortTitle{Ionized gas emission}

\author[1,2]{Greta Guidi\orcidlink{0000-0002-7002-8928}}
\ShortName{Guidi et al.} 
\author[3,4]{Christian Rab\orcidlink{0000-0003-1817-6576}}
\author[3]{Barbara Ercolano\orcidlink{0000-0001-7868-2740}}
\author[3]{Michael L. Weber\orcidlink{0000-0002-4983-0422}}
\author[5]{Claudio Codella\orcidlink{0000-0003-1514-3074}}
\author[6]{Izaskun Jim\'enez-Serra\orcidlink{0000-0003-4493-8714}}
\author[7,8]{Evgenia Koumpia\orcidlink{0000-0002-1922-8692}}
\author[9]{John D.\ Ilee\orcidlink{0000-0003-1008-1142}}
\author[10]{Enrique Mac{\'\i}as\orcidlink{0000-0003-1283-6262}}
\author[10]{Elena Viscardi\orcidlink{0009-0005-5187-6074}}
\author[11,12]{Yinhao Wu \begin{CJK*}{UTF8}{gbsn}(吴寅昊)\end{CJK*} \orcidlink{0000-0003-3728-8231}}
\author[5]{Francesca Bacciotti\orcidlink{0000-0001-5776-9476}}
\author[13]{Asmita Bhandare\orcidlink{0000-0002-1197-3946}}
\author[1,5]{Eleonora Bianchi\orcidlink{0000-0001-9249-7082}}
\author[14]{Tyler Bourke\orcidlink{0000-0001-7491-0048}}
\author[8]{Luca Cacciapuoti\orcidlink{0000-0001-8266-0894}}
\author[2]{Antonio Garufi\orcidlink{0000-0002-4266-0643}}
\author[1]{Geoffroy Lesur\orcidlink{0000-0002-8896-9435}}
\author[15]{Vincent Pi\'etu\orcidlink{0009-0006-3497-397X}}
\author[5]{Linda Podio\orcidlink{0000-0003-2733-5372}}
\author[5]{Giovanni Sabatini\orcidlink{0000-0002-6428-9806}}
\author[16]{Leonardo Testi\orcidlink{0000-0003-1859-3070}}
\author[10,17,5]{Claudia Toci\orcidlink{0000-0002-6958-4986}}

\affiliation[1]{Univ. Grenoble Alpes, CNRS, IPAG, 38000 Grenoble, France}
\emailAdd{greta.guidi@univ-grenoble-alpes.fr}
\affiliation[2]{INAF, Istituto di Radioastronomia, Via Gobetti 101, I-40129, Bologna, Italy}
\affiliation[3]{University Observatory, Faculty of Physics, Ludwig-Maximilians-Universit\"at M\"unchen, Scheinerstr. 1, 81679 Munich, Germany}
\affiliation[4]{Max-Planck-Institut f\"ur extraterrestrische Physik, Giessenbachstrasse 1, 85748 Garching, Germany}
\affiliation[5]{INAF, Osservatorio Astrofisico di Arcetri, Largo E. Fermi 5, I-50125, Firenze, Italy}
\affiliation[6]{Centro de Astrobiologia (CAB), CSIC-INTA, Carretera de Ajalvir km 4, Torrej\'on de Ardoz, 28850, Madrid, Spain}
\affiliation[7]{Joint ALMA Observatory, Alonso de Cordova 3107, Vitacura, Santiago, Chile}
\affiliation[8]{European Southern Observatory, Alonso de Cordova 3107, Vitacura, Santiago, Chile}
\affiliation[9]{School of Physics and Astronomy, University of Leeds, Leeds, UK, LS2 9JT, UK}
\affiliation[10]{European Southern Observatory, Karl-Schwarzschild-Strasse 2, D-85748 Garching bei M\"unchen, Germany}
\affiliation[11]{Shanghai Astronomical Observatory, Chinese Academy of Sciences, Shanghai 200030, People's Republic of China}
\affiliation[12]{School of Physics and Astronomy, University of Leicester, Leicester LE1 7RH, UK}
\affiliation[13]{Universit\"{a}ts-Sternwarte, Fakult\"{a}t f\"{u}r Physik, Ludwig-Maximilians-Universit\"{a}t M\"{u}nchen, Scheinerstr. 1, 81679 M\"{u}nchen,Germany}
\affiliation[14]{SKA Observatory, Jodrell Bank, Lower Withington, Macclesfield, SK11 9FT, UK}
\affiliation[15]{IRAM, 300 rue de la piscine, F-38406 Saint Martin d'H\`eres, France}
\affiliation[16]{Alma Mater Studiorum Universit\`a di Bologna, Dipartimento di Fisica e Astronomia (DIFA), Via Gobetti 93/2, 40129 Bologna, Italy}
\affiliation[17]{Departamento de Fisica aplicada III, ETSI Universidad de Sevilla,  Camino de los Descubrimientos, 41092 Sevilla}

\abstract{Protoplanetary disks represent a crucial stage in the evolution of Young Stellar Objects towards the formation of fully formed planetary systems.
While substantial progress has been made in the last decades in the characterization of the dust and molecular gas in these systems, the ionized component remains poorly understood. 
Ionized gas traces important processes such as photoevaporation, accretion, disk winds, and jets, and therefore is key to studying disk dynamics, evolution, and ultimately planet formation. 
In this paper, we investigate the capabilities of the forthcoming SKA telescope to probe this component in protoplanetary disks within nearby star forming regions.
We present state-of-the-art simulations of photoevaporative, magneto-thermal, and magnetohydrodynamic winds, and generate theoretical predictions and synthetic SKAO observations to assess its potential in detecting and characterizing free--free emission and Hydrogen recombination lines. 
Finally, we discuss synergies with complementary facilities and how they will provide a comprehensive, multi-scale view of disk winds and offer critical insights on the mechanisms driving disk evolution and the onset of planet formation. 
}

\begin{document}
\maketitle

\section{Introduction}
Protoplanetary disks of gas and dust are formed as an outcome of angular momentum conservation during the collapse of molecular cloud cores, and regulate the accretion of material onto the forming star \citep{StahlerPalla2004, Zhao2020}. 
A number of complex physical and chemical processes shape the evolution and lifetime of disks, with direct consequences on the timescales and initial conditions for planet formation \citep{Lesur2023}.
Understanding the mechanisms regulating disk dispersal, mass loss, and angular momentum transport is crucial for connecting the physical properties of protoplanetary disks to the formation and architecture of planetary systems. 

\begin{figure}[h!]
    \centering
    \includegraphics[width=0.9\linewidth]{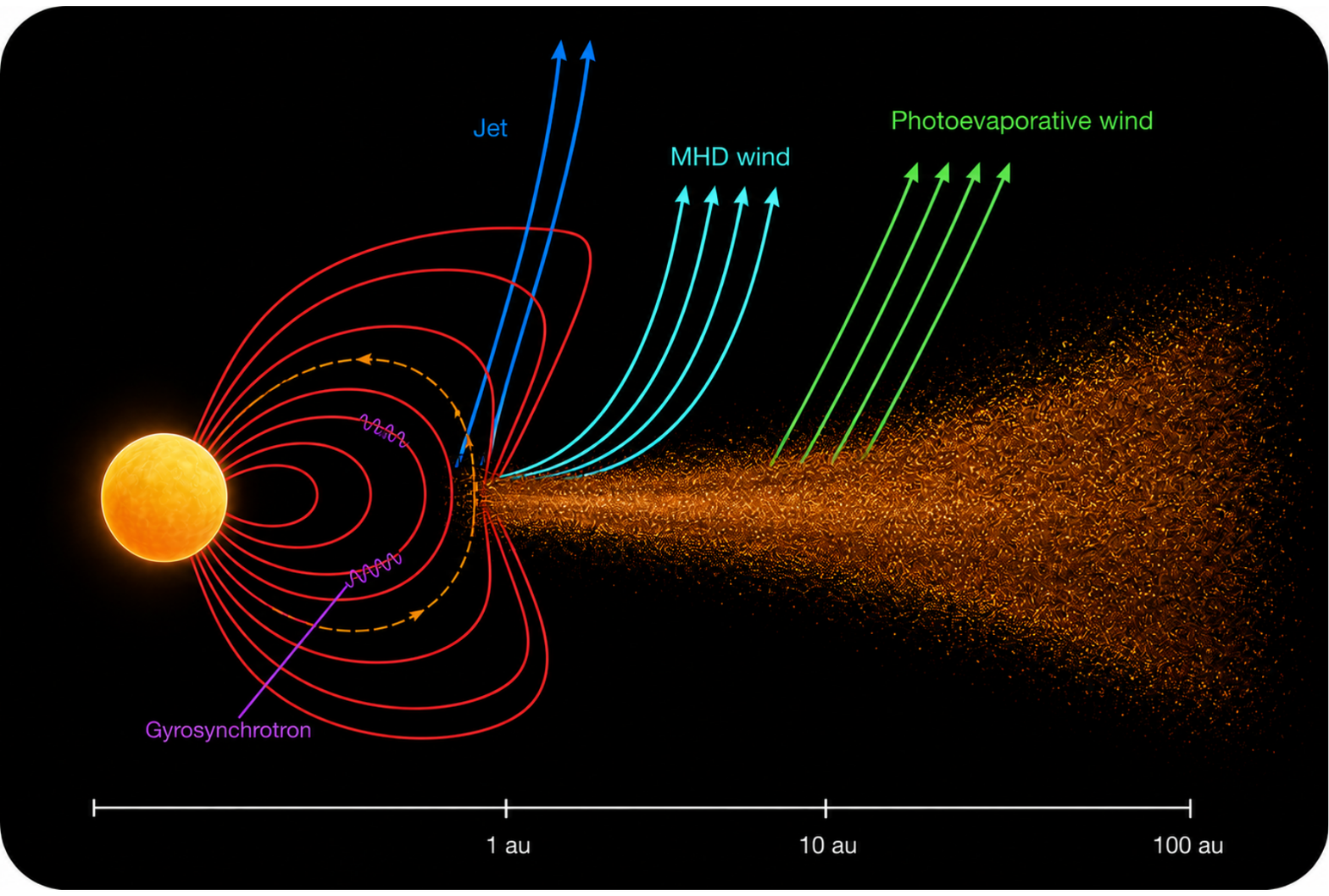}
    \caption{Schematic representation of a protoplanetary disk highlighting accretion and ejection processes. The spatial scales at the bottom are indicative only and serve to illustrate the approximate emitting regions of the different components.}
    \label{fig:disksection}
\end{figure}

While significant advances have been made in characterizing the dust and molecular gas components of disks -- particularly thanks to facilities such as ALMA (Atacama Large Millimeter/submillimeter Array) and VLT (Very Large Telescope) -- the ionized component remains comparatively less explored. This is typically observed via atomic lines in the optical and infrared range, hydrogen recombination lines and free--free continuum emission, although these tracers are often faint and difficult to isolate, limiting our ability to fully constrain the properties and origin of the ionized gas. 
Ionized gas is associated with a range of key physical processes, including accretion, photoevaporation, disk winds, and jets (see Figure \ref{fig:disksection} for a schematic illustration). These processes play a critical role in regulating the lifetime and structure of the disk, as well as in determining the mass available for planet formation. 
Observing and interpreting the emission from ionized gas provides insight into the energy input from the central star, the mechanisms of disk dispersal, and the interface between disks and stellar magnetospheres.


Observational evidence indicates that disk lifetimes are of the order of a few million years, with disk dispersal and decay of stellar accretion rate occurring on similar timescales \citep{Mamajek2009,Fedele2010}. While the exact mechanisms responsible for the loss of disk material are still under debate, the current paradigm identifies photoevaporative winds and MHD (Magnetohydrodynamics) winds as the main drivers of disk dispersal \citep{ErcolanoPascucci2017}. The latest observational evidence suggests MHD winds as dominant at the early stages and photoevaporative winds taking over during the final phases of disk clearing \citep{Pascucci2023}.
Slow molecular outflows have been detected in a large number of sources, across different evolutionary stages. 
At the Class~II stage, the typical tracers of disk winds are forbidden lines such as H$_2$, [OI] and [SII], associated with a low velocity component (typically $\lesssim$10~km/s). Thanks to ALMA, such slow winds are detected also in molecular CO \citep{Guedel2018}. 
The ejection rates estimated for these outflows are comparable to the stellar accretion rate, and - when measurements of both components are available - $\dot M_{\mathrm{wind}}$ is approximately 50 $\times$ $\dot M_{\mathrm{jet}}$ \citep[see][and references therein]{Pascucci2023}. Although current statistics remain limited, this finding suggests that disk winds may constitute the primary source of mass loss, dominating over jets.

The Square Kilometre Array (SKA), with its unprecedented sensitivity at centimeter wavelengths, will open an additional window to detect this ionized component in such a crucial stage of evolution. 
In particular, free--free emission detected at high spatial resolution and the capability to observe radio recombination lines offers a new and powerful tool to investigate disk dissipation mechanisms and processes in the inner regions. 

In this paper we investigate the capabilities of SKA AA4 \citep{braun2019anticipatedperformancesquarekilometre} in unveiling ionized gas tracers in disks. We outline the key science questions that can be addressed and highlight the most promising observational signatures by generating simulated observations from theoretical predictions of MHD and photoevaporative winds.

\section{Disk wind theory}
Disk winds are widely recognised as a fundamental agent of protoplanetary disk evolution, regulating the redistribution and eventual dispersal of disk material, and thereby shaping the environment for planet formation. Two broad classes of winds have been identified: those driven primarily by magnetic stresses (MHD winds), and those launched thermally by high-energy irradiation from the central star (photoevaporative winds). While both processes may operate simultaneously, their physical origins, mass fluxes, and evolutionary signatures differ in ways that remain the subject of active debate. 
A closely related phenomenon is that of jets, consisting of highly collimated outflows of material launched from regions very close to the protostar \citep[see][and references therein]{Pudritz2007}. They are found to be more prominent in early--stage YSOs, but they are observed throughout all evolutionary phases and across the full range of protostellar masses (e.g., \citealt{Anglada2018}). Although the exact origin and acceleration mechanisms of jets are not fully established, their strong collimation suggests a magnetic origin and a direct connection to the same magnetohydrodynamic processes generating magnetically driven disk winds, discussed in this section. For a comprehensive discussion of jets and their observational prospects with the SKAO, we refer to the chapter by \citet{Sabatini01.2026.SKA}. 

MHD winds arise when magnetic fields threading the disk extract angular momentum and energy, enabling mass to be centrifugally flung away along open field lines \citep{Blandford1982,PudritzNorman1983,BaiStone2013}. The efficiency of this mechanism depends sensitively on the ionisation state and magnetic diffusivities within the weakly ionised disk surface layers, with non-ideal MHD effects such as the Hall term playing a decisive role in recent global simulations \citep[e.g.][]{Sarafidou2024}. MHD-driven mass loss is predicted to reach rates of $\sim 10^{-8} - 10^{-7}\,M_\odot\,\mathrm{yr}^{-1}$ in typical T Tauri disks, and importantly, it simultaneously drives disk accretion by transporting angular momentum vertically, rather than radially, through the disk \citep{Bai2016,Lesur2021}.

Photoevaporative winds, by contrast, are powered by heating of the upper disk atmosphere by stellar extreme ultraviolet (EUV), far-ultraviolet (FUV), and X-ray photons \citep{Hollenbach1994,Alexander2006, Ercolano2018}. 
Once the gas is heated above the local escape temperature, a hydrodynamic outflow is launched \citep{Owen2010}. Early models found characteristic mass-loss rates of order $10^{-10} - 10^{-8}\,M_\odot\,\mathrm{yr}^{-1}$, depending on the irradiation spectrum. More recent work has demonstrated the importance of improved radiative transfer and thermal-chemical coupling: for example, \citet{Picogna2019,Picogna2021,Ercolano2021,Ercolano2022} refined the initial models, including dependencies on stellar mass and irradiating spectrum. Most recently, \citet{Sellek2024} showed  that enhanced cooling efficiencies in the outflowing gas can significantly reduce X-ray-driven mass-loss rates relative to earlier estimates, bringing them more in agreement with observations. The interplay between photoevaporation and disk substructures, including dead zones and planet-induced gaps, further modulates the wind properties and observable signatures \citep{Garate2021,WeberM2024}.

The relative contributions of MHD and photoevaporative winds remain uncertain and are likely time-dependent. MHD winds may dominate early, when strong magnetic flux is retained, while photoevaporation is expected to play a crucial role in the final dispersal of the disk \citep{ErcolanoPascucci2017, Pascucci2023}. Importantly, both processes have direct implications for planetesimal and planet formation. Disk winds remove gas mass and angular momentum, altering the pressure gradients that regulate dust evolution and pebble drift. This can accelerate planetesimal formation via streaming instabilities, while at the same time reducing the gas reservoir available for giant planet formation. The redistribution of gas by winds may also influence planetary migration pathways \citep{Monsch2023,WuChen2025}. Thus, disk winds are not only a dispersal mechanism but also an active participant in sculpting planetary system architectures.

In summary, the theory of disk winds has advanced considerably over the past decade, aided by multi-physics simulations and improved stellar irradiation models \citep{Mori2025}. 
Some of the most recent studies consider planet-disk interaction with 3D non-ideal MHD disk winds \citep{AoyamaBai2023,Wafflard-FernandezLesur2023,Wafflard-FernandezLesur2025}, multi-fluid simulations with MHD disk winds \citep{Elbakyan2022,Wu2023,Wu2024}, and MHD disk winds with dust growth \citep{KawasakiMachida2025}. 
Yet, significant open questions remain: how MHD and thermal winds coexist and compete, how their mass fluxes scale with stellar properties, and how these winds ultimately dictate the transition from gas-rich disks to planetary systems. Tracing the ionised component of these winds might offer some valuable insights into these questions. 
Recent results from optical/near-infrared observations of OI/Ne in PPDs (Protoplanetary Disks) with HST and JWST \citep[e.g.][]{Fang2018, Campbell-White2023, Bajaj2024}
are addressing the issue of probing the presence of disk winds (see also Section \ref{sec:synJWST}), although the results so far remain inconclusive. 

To illustrate how SKAO future observations of ionized gas emission could help in this direction, we present in Section \ref{sec:models} a set of simulations of photoevaporative and MHD winds along with the corresponding predictions and detection limits for SKAO.

\section{Radio emission from protoplanetary disks}
\label{sec:radioemission}
The radio window provides access to a variety of emission processes in the surroundings of protoplanetary disks: at these long wavelengths the emission from the material surrounding the protostar dominates the SED (Spectral Energy Distribution) of Young Stellar Objects. 
Line emission is detected from rotational transitions from the disk molecular layers \citep[see chapter by][]{Podio01.2026.SKA}, and the continuum traces both the solid and the ionized gas component. 
In particular, dust grains are the primary source of continuum emission from the infrared range (where stellar radiation scattered by grains in the disk surface prevails) to the millimeter/centimeter regime (where thermal emission from re-processed radiation is produced by larger grains). 
At centimeter wavelengths, additional mechanisms become increasingly important in the total flux, most notably free--free emission from ionized gas in the star surroundings, i.e. from collimated jets or outflows, accretion shocks near the forming star, and disk winds.
Another significant contributor can be non-thermal gyrosynchrotron emission from magnetically active stars.
Anomalous Microwave Emission (AME) is another potential mechanism that may contribute to the centimeter-wavelength continuum from protoplanetary disks. Defined as an excess in Galactic radiation between 10 and 60 GHz, AME is most commonly attributed to rapidly rotating nanometer-sized dust grains \citep[see][for a review]{Dickinson2018}. This mechanism has been suggested to explain the SED in a few disks \citep{Greaves2018, Baobab2024}, but separating its contribution from other emission components remains challenging. 
Further details on this component and its detectability with SKAO are provided in the chapter by \citet{Vidal01.2026.SKA}.  

A common method used to distinguish between the different processes is to measure the cm spectral slope.  Free--free emission coming from disk atmospheres is expected to have a spectral index $\alpha$ between 
-0.1 and 2 for the optically thin and optically thick case respectively \citep{Ubach2017}. Gyrosynchrotron emission from corona flares are predicted to have slopes around 2 in the optically thick side and steeply negative slopes, up to -15, in the optically thin side \citep{Guedel2002,Golay2023}. 
Surveys of nearby star-forming regions have shown a trend of decreasing spectral index of the cm-emission with evolutionary stage: this has been interpreted as free--free emission being the dominant mechanism in early-type YSOs (Class 0 to Class II, becoming increasingly optically thin by the Class II stage), and gyrosynchrotron prevailing in more evolved Class III sources \citep{Dzib2013,Liu2014,Coutens2019}. While there is evidence of magnetic activity in classical T Tauri stars, gyrosychrotron emission can be absorbed by the circumstellar winds at the stages where they are dominant, preventing its detection at radio wavelengths \citep{Guedel2002}. 

Multiple studies of nearby star-forming regions have shown that excess emission beyond the thermal dust continuum becomes significant, and in a few cases even dominant, already at millimeter wavelengths. \citet{Rodmann2006} reported that approximately 20\% of the 7~mm flux arises from free--free emission, with the remaining 80\% attributed to dust, in line with more recent findings by \citet{Garufi2025}. A broader range of non-dust contributions was observed by \citet{Ubach2017} in T Tauri stars within the Chamaeleon and Lupus regions, where the fraction of non-dust emission at 7~mm varies from just a few percent up to 80\%. \citet{Coutens2019} also found that less than 30\% of the detected continuum emission at 10 GHz (3\,cm) can be attributed to dust, and hence the remaining 70\% is due to either free--free or to synchrotron emission. This wide spread suggests a mix of underlying mechanisms, potentially including non-thermal processes from some sources displaying rapid variability on timescales of minutes.

Estimates of this excess emission have often relied on extrapolating the non-dust contribution from power-law fits to the cm-wavelength integrated fluxes,  or vice versa, estimating the dust component from the mm-integrated fluxes. However, this approach has limitations due to the fact that in either cases the SED can deviate from a simple power-law \citep[e.g.][]{Guidi2022}. 
Recently, \citet{Painter2025} addressed the non-power-law nature of the SED by adopting multi-component, empirical model prescriptions to disentangle the dust and other emission mechanisms of eight young systems in the Taurus region. 
Their inferred dust disk spectra all show significant curvature making single power-law modelling uncertain. While they found no evidence for synchrotron or spinning dust grain emission, all of their targets exhibited free--free emission contributing 5-50\% of the 43\,GHz fluxes. 
Spatial resolution can help directly disentangling dust thermal emission from other types of contributions: for instance, \citet{Macias2016} distinguished two components of the 3~cm free--free emission in GM~Aur: one arising from an ionized radio jet perpendicular to the disk, and another consistent with a photoevaporative wind driven by EUV radiation. \citet{Guidi2022} partially resolved the 3.6~cm emission in HD~163296, finding a more compact non-dust emission coming within a radius of approximately 5~au and with a fraction contribution of 40\% of the total flux. In this case the derived spectral index at centimeter wavelengths, around 0.1, was consistent with optically thin free--free emission produced by a disk wind.

\subsection{Line observations}
At millimeter wavelengths, high-resolution observations of protostellar objects (10$^4$--10$^5$ yr old) conducted with ALMA have revealed the presence of relatively slow molecular emission, characterized by velocities on the order of approximately 10 km s$^{-1}$, originating from an extended region of the protoplanetary disk, and reaching radial distances of about 40 to 50 au. These detections, reported in studies by \citet{Tabone2017,Tabone2020}, \citet{Lee2018}, \citet{Nazari2024}, and \citet{DeSimone2024}, were made possible through the identification of emission lines from molecules such as sulfur monoxide (SO), sulfur dioxide (SO$_2$), methanol (CH$_3$OH), and silicon monoxide (SiO). 
Collectively, these findings provide compelling and increasingly robust evidence that the observed disk winds are chemically stratified and transition from an ionized phase to a molecular phase that is rich in a wide variety of molecules. 
This phenomenon is intimately linked to a gas phase that is chemically enriched, suggesting active chemical processing and potentially significant implications for disk evolution and planet formation.

 


\subsection{Diagnostic tracers and current limitations}
We summarize here the main diagnostic tracers than can be employed to disentangle between the different mechanisms generating continuum emission at radio wavelengths. 

\textit{Spectral index}: the flux spectral index at cm-wavelengths is one of the primary and most accessible diagnostic for disentangling between dust emission ($\alpha >$2), free--free emission (-0.1$\leq \alpha \leq$2, depending on optical depth), non-thermal synchrotron (typically from jets) or gyrosynchrotron (from stellar magnetosphere) emission, generally exhibiting a negative spectral index. 
Observations at cm-wavelength of protoplanetary disks are often sparse, and do not allow for a detailed spectral characterization. More recent studies sampling broad frequency ranges are highlighting a more complex shape of the SED than a single power-law, suggesting that multiple mechanisms and/or physical regimes are at play \citep[e.g.][]{Hashimoto2025, Chung2025}. This highlights the need for high sensitivity broadband observations, to accurately model the various mechanisms at work. 

\textit{Morphology}: spatially resolving the cm-continuum is an extremely powerful tool to locate the emitting regions of the different components (see schematic picture in Figure \ref{fig:disksection}). While dust emission can extend to tens of au, ionized gas emission originates closer to the star and can have distinctive emission regions. For the case of disk winds, we show in Section \ref{sec:models}, free--free and radio recombination lines (RRLs) emission present different spatial extent and morphologies depending on the wind model.  

\textit{Time variability:}
Time variability has been  detected in mm and cm emission: for example, \citet{Ubach2017} found that most of the target T--Tauri stars in Chamaeleon and Lupus star-forming regions displayed a variability at 7mm from about 30\% up to a factor of 3 across timescales from months to years. Several disks show spectral indices consistent with grain growth to cm-size ``pebbles''. 
Interestingly, the most radio-active sources (largest excess/variability) tend to have less evidence for large grains, suggesting that strong free--free/non-thermal emission can coincide with limited grain growth signatures. 
Similarly, \citet{Hashimoto2025} observed a variable spectral index at cm wavelengths (with timescales of the order of days) in the Class I source WL~17, and fit the observations using two independent free--free components that they associate with different knots of ionized gas, likely coming from a jet. 
Non thermal processes such as chromospheric or magnetospheric emission from the central star typically present a variability on shorter timescales, ranging from minutes to days, and larger flux variations by factors of a few.

\textit{Polarization:} At (sub-)millimetre wavelengths, disk emission can show low ($\lesssim$1-2\%) linear polarization produced by dust self-scattering or grain alignment to the disk radiation field \citep{Kataoka2016, Tazaki2017, Yang2016, Bacciotti2018}, or to the disk magnetic field \citep{Ohashi2025}. 
At longer wavelengths (few centimetres), polarization measurements can instead become sensitive to the presence of jets. While free--free emission is essentially unpolarized, non-thermal synchrotron radiation can exhibit strong linear polarization ($\sim$10-30\%) tracing magnetic fields \citep{Carrasco-Gonzalez2010,Rodriguez-Kamenetzky2017}. 
Circular polarization generated by Gyrosynchrotron emission from accelerated electrons can display an intermediate degree of circular polarization \citep{Guedel2002}, while coherent processes such as plasma emission or cyclotron masers can give rise to a very high degree of circular polarization 
\citep[e.g. T Tau,][]{Loinard2007}. 
Although rare, both circular and linear polarization has been observed at cm emission from TTauri stars \citep{Phillips1996} and  connected to stellar magnetic flares. 
Polarimetric observations across the mm-cm regime are therefore a powerful tool for distinguishing between thermal and non-thermal contributions to the continuum emission. 


\section{Theoretical predictions from simulations of MHD and photoevaporative winds}
\label{sec:models}
Considerable efforts have been made to estimate the signatures of disk winds on continuum and line tracers, from optical to radio wavelengths \citep[e.g.][]{ercolano2010,Weber2020,Ricci2021,nakatani2025}. 
In the cm range, a major advantage is that the continuum emission from electrons in the ionized winds dominates over the dust continuum \citep[see previous Section~\ref{sec:radioemission} and the chapter by][]{Garufi01.2026.SKA}. 
For RRLs such as the Hydrogen alpha transitions (i.e. transitions to a contiguous principal quantum number), the intensity decreases with increasing wavelength but the line/continuum contrast improves, making the cm-range a favorable regime with ratios of 1-10\% \citep[see][for line emissions in photoevaporative winds]{Pascucci2012}. 
A characterization of such recombination lines is often considered a key diagnostic to probe the presence of disk winds. Specifically, line profiles and velocities can unveil the unbound nature of the gas emission, pointing to material decoupled from the disk rotation. 
Furthermore, the shape and width of the line are expected to be quite different in photoevaporative winds and MHD-driven winds, with the latter being faster \citep[$\ge$ 40 km/s, see e.g.][and this work, Section \ref{sec:models}]{Fang2018}. 

In this Section we present predictions of free--free and hydrogen recombination lines emission arising from both photoevaporative and magnetically-driven disk winds. Our goal is to demonstrate the detection capabilities of SKA-Mid AA4 array and assess the potential to disentangle the origin of the ionized gas using both line and continuum tracers.

\subsection{Model description and results}
\label{sec:modeldescription}
We use existing simulations from various disk wind models, including pure photo-evaporative (PE, e.g. \citealt{Picogna2019,Weber2020}, PLUTO code\footnote{PLUTO: \url{https://plutocode.ph.unito.it/}}), magneto-thermal (MT, \citealt{Sarafidou2024,weber_thermal_2025}, NIRVANA code\footnote{NIRVANA: \url{https://gitlab.aip.de/ziegler/NIRVANA}}), and isothermal magneto-hydro-dynamics wind models (MHD, \citealt{Lesur2021}, semi-analytic\footnote{see also \url{https://github.com/glesur/PPDwind}}). These models are used to produce synthetic observables for the free--free emission and for the Hydrogen recombination lines (\holine). 

For the free--free emission, we post-process the (magneto)-hydro-dynamic models with the radiative transfer and X-ray enabled ionisation code \mocassin \citep{ercolano2003, ercolano2005, ercolano2008}. The radiation thermochemical code \prodimo (PROtoplanetary DIsk MOdel\footnote{\prodimo git-rev: 2e42470c, \url{https://prodimo.iwf.oeaw.ac.at/}}, \citealt{woitke_radiation_2009,Kamp2011,thi_radiation_2011}) is used for the \holine. We follow an already well-tested approach used for other wind-tracing observables such as \oiline, \Neline, and the \htline lines \citep[e.g.][]{Weber2020,rab2023,weber_thermal_2025}. In the following, we only describe specifics relevant to the SKA observables and some relevant differences in the models.
We refer the reader to the above-mentioned papers for additional details concerning the modelling approach. 

\subsubsection{Modelling approach}
In all the presented models the central star is representative of a classical T Tauri star with $L_\mathrm{*}=1\,\mathrm{L_\odot}$, $M=1\,\mathrm{M_\odot}$, $L_\mathrm{X}\approx2\times 10^{30}\,\mathrm{erg\,s^{-1}}$, and a primordial disk (i.e. no substructure or gaps). However, mainly due to computational limitations, not all models have the exact same underlying disk structure. 

The computational domain for the disk of the MT models is limited to $r=0.5-15\,\mathrm{au}$, $r$ being the distance from the central star. The disk in the PE and MHD models extends from $r=0.3-60\,\mathrm{au}$. 
Furthermore, the MHD models have a shallower radial surface density profile compared to the other models; this is a consequence of how the semi-analytic models of \citet{Lesur2021} are constructed. These differences somewhat limit a direct comparison of the models; however, they still give a general picture of what impact the different wind-driving mechanisms and modelling approaches have on observables relevant for SKAO.

Furthermore, parameters such as the magnetic field strength are not relevant for PE models, and on the other hand, the MHD disk-wind model of \cite{Lesur2021} assumes an isothermal disk; hence, the wind solution is independent of the stellar properties. For the post-processing, we place the disk at a distance of $140\,\mathrm{pc}$, and test disk inclinations of $i=[0^\circ,20^\circ,40^\circ,60^\circ,80^\circ]$ degrees. For the spectral lines, we use an intrinsic spectral resolution of $\delta v=0.25\,\mathrm{km/s}$, which is well above the achievable spectral resolution of SKA-Mid at Band 5, corresponding to $\sim$ 10\,m/s.

\paragraph{Free--free emission:}

The prediction of free--free emission employs the improved photon packet statistics introduced by \citet{Ricci2021} to minimize Monte Carlo noise. For both the PE and MT models, we adopt the same illuminating spectrum composed of two components: a soft X-ray component calculated by \citet{ercolano2009} with a luminosity of $2\times10^{30}$ erg s$^{-1}$, and a blackbody component with a temperature of 12000 K and a bolometric luminosity of $L_\mathrm{accr}=2.6\times10^{-2} \,\mathrm{L_\odot}$. The latter represents emission associated with accretion.

\paragraph{Hydrogen Recombination lines:}

The excitation model for Hydrogen in \prodimo is based on data from the NIST database \citep{kramida_critical_2010,NIST_ASD} and so far only considered energy levels up to $n=25$ (i.e. near/mid infrared lines). The details of this implementation and an application to Hydrogen emission lines from disks around massive stars can be found in \citet{backs_massive_2023}. For the purpose of this paper, we extended this model up to $n=200$, including in total $19,900$ transitions. For the missing data (e.g. Einstein-A coefficients) we applied the RADZ1 code from \citet{storey_fast_1991}.
We note that for the line-predictions we also include models with different L$_\mathrm{accr}$, as the UV luminosity is most relevant for the chemistry and line emission.

\subsubsection{Model results}
\label{sec:modelresults}
For the free--free emission, we present the proper SKA simulations in Sec. \ref{sec:skasimulations}. For the \holine, we discuss here the theoretical model results in more detail. In contrast to the free--free emission, the \holine will also provide information on the velocity structure of the wind, with the limitations that SKA might not be sensitive enough to (always) detect those lines.   

\begin{figure}
    \centering
    \includegraphics[width=0.32\linewidth]{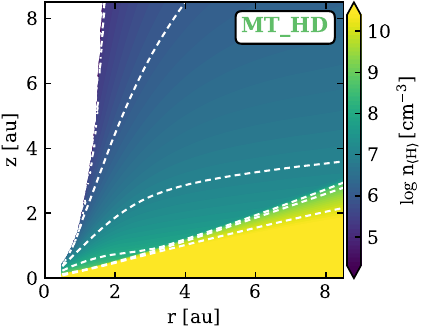}
    \includegraphics[width=0.33\linewidth]{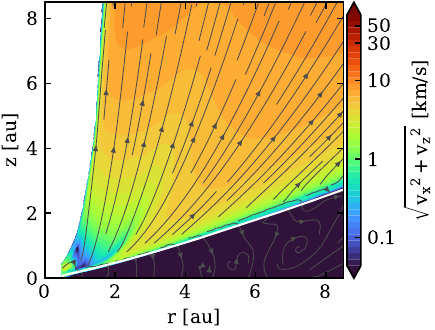}
    \includegraphics[width=0.31\linewidth]{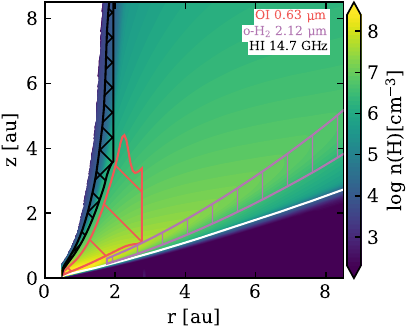}\\
    \includegraphics[width=0.32\linewidth]{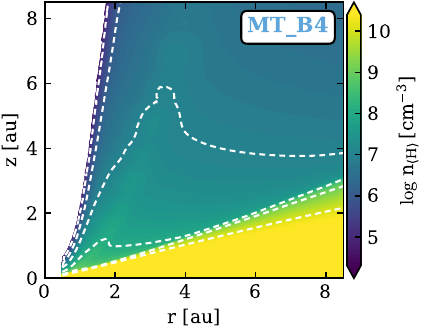}
    \includegraphics[width=0.33\linewidth]{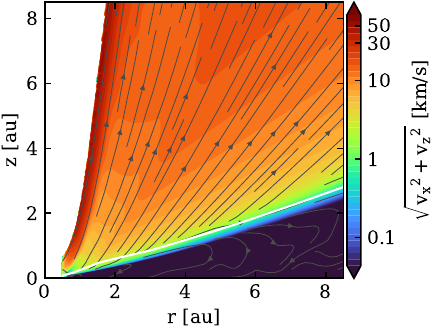}
    \includegraphics[width=0.31\linewidth]{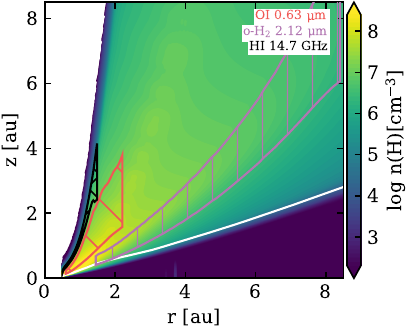}   
    \caption{Two-dimensional density structure (total hydrogen number density), wind velocity field, and main line emitting regions (boxes), with the neutral hydrogen number density in the background. The dashed white lines in the left panels mark the contour levels indicated in the colourbar. The white solid lines in the middle and right panels indicate a visual extinction of unity.  The top row shows the MT\_HD wind model (i.e., no magnetic fields), the bottom row shows the MT\_B4 magneto-thermal wind model. For easier comparison, we use the same $L_\mathrm{accr}\approx0.3\,\mathrm{L_\odot}$ in both models here. The MT\_B4 model has a faster and denser wind compared to the MT\_HD model. As a consequence, the radial optical depths are higher, which results in slightly more compact \holine and \oiline emission, but in a higher and more prominent H$_2$ emission layer (i.e., the wind is molecular).}    
    \label{fig:magnetothermal}
\end{figure}

At first, we discuss theoretical results from the models regarding the velocity structure and the density of the wind. The MT models are, physically speaking, the most self-consistent ones, as they include both the thermal and magnetically driven launching mechanisms, which may be the most realistic scenario \citep[see e.g.][]{Bai2016}. In Fig.~\ref{fig:magnetothermal}, we show the disk structure, the wind velocity field, and the main line-emitting regions. We consider here three lines, the \holine at 14.7 GHz, the \htline, and the commonly used wind tracer \oiline. We note that within a model, all \holine lines in the 5b band of SKA-Mid considered here are very similar in their properties, in particular, the line-width and general shape of the line profile.

We show one model with magnetic fields and thermal-to-magnetic pressure ratio of $\beta=10^{4}$ (model MT\_B4), and one without magnetic fields (thermally driven, MT\_HD) but still using the same underlying modelling approach. The MT\_B4 model shows a denser and faster wind, which results in mostly compact line-emitting regions for the \holine, which sits just above the \oiline emitting region. The picture is similar in the MT\_HD model, but the line-emitting regions are vertically more extended. The comparison of the \htline emission in the two models nicely shows the impact of the higher wind density. In a denser wind, the far-UV radiation cannot penetrate as deep into the wind/disk, hence molecular hydrogen also survives higher up in the wind, and traces higher velocities compared to the \htline in the MT\_HD model \citep[see also][]{rab2022,Sellek2024,weber_thermal_2025}. As a result, the extent of the pure atomic gas will be more compact, also resulting in more compact line emission. 

\begin{figure}
    \centering
    \includegraphics[width=0.32\linewidth]{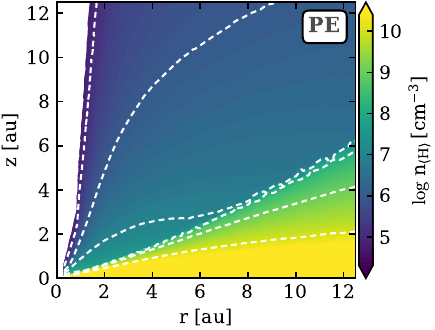}
    \includegraphics[width=0.33\linewidth]{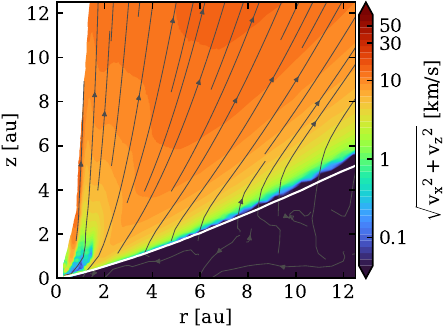}   
    \includegraphics[width=0.31\linewidth]{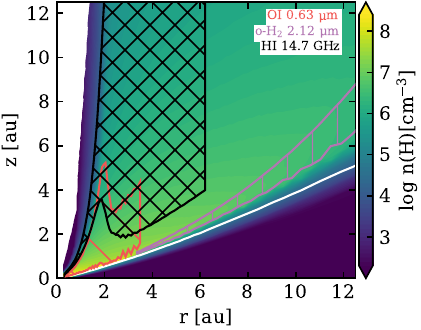}\\
    \includegraphics[width=0.32\linewidth]{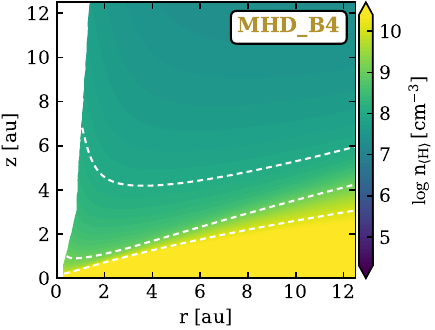}
    \includegraphics[width=0.33\linewidth]{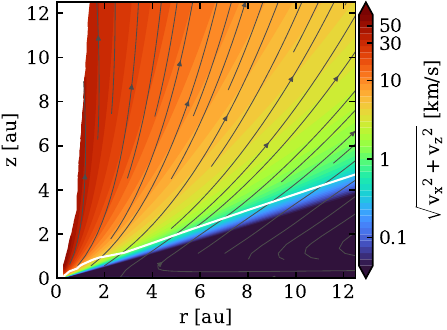}
    \includegraphics[width=0.31\linewidth]{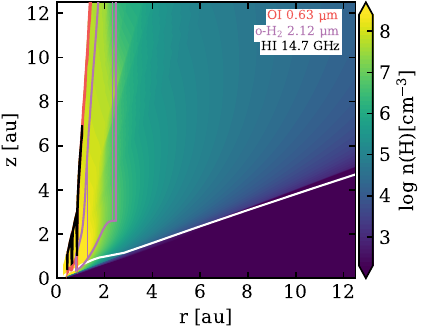}\\    
    \caption{Same as Fig.~\ref{fig:magnetothermal} but for the PE (top row) and the MHD\_B4 model. These two models show nicely the potentially drastic differences between the two wind-launching mechanisms. The MHD model predicts a very dense and fast wind, especially in the innermost disk region and a strong radial gradient in the wind velocities. Whereas, the PE model predicts low (close to zero) velocity wind in the inner region with a roughly constant wind velocity further out.}
    \label{fig:MHDPE}
\end{figure}

In Fig.~\ref{fig:MHDPE}, we show the same plots as in Fig.~\ref{fig:magnetothermal} for the two other model types, PE and MHD. As already noted, the computational domain is more extended in those models, compared to the MT models, and hence we do not directly compare them. The smaller radial extent of the MT models is likely not an issue, as the emitting regions of the lines are restricted to the inner 10~au in all our models. However, the larger inner radius in the MT models might have an impact on the line profiles and fluxes. In particular for magnetically-driven winds, this inner region might be the origin of a high-density and high-velocity flow.%

In general, the differences between a pure thermally-driven wind and a (thermo-)magnetically-driven wind are similar to what was discussed already for the MT models. However, the differences are much more pronounced. Interestingly, the PE model now shows a radially more extended \holine emitting region, which is likely a result of the slightly lower wind density, compared to the MT\_HD model. For the model MHD\_B4, which also uses $\beta=10^4$, we now see very compact emission regions for all three lines. This is a result of the high density in the wind, and hence the outer disk is well shielded from any radiation. For such a case, all three lines considered only trace the innermost, high-velocity part of the disk wind, as most of the wind is molecular, and hence only the walls of the outflow cavity are visible in line emission.%
\begin{figure}
    \centering
    \includegraphics[width=0.49\linewidth]{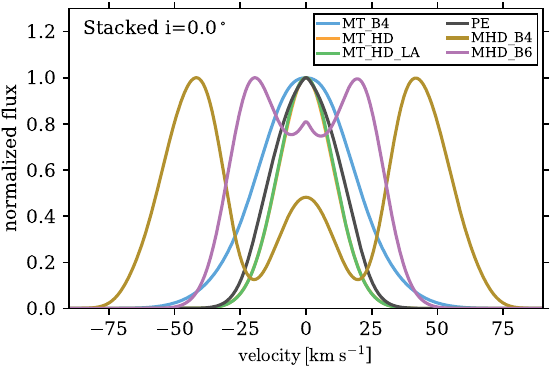}
    \includegraphics[width=0.49\linewidth]{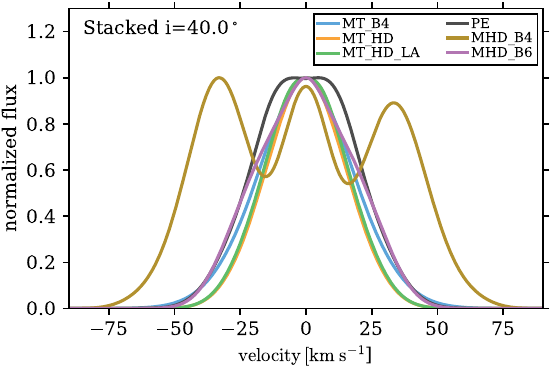}
    \caption{Stacked and peak normalised \holine line profiles for the various models. The left panel shows the profiles for the disk seen face-on, the right panel shows the profiles for an inclination of $i=40^\circ$ (right panel). Please note that in the left panel, the model MT\_HD (orange line) is barely visible, as it is almost identical with the MT\_HD\_LA (green line) model.}
    \label{fig:profiles}
\end{figure}

Before we continue with detailing simulations specifically for SKAO (Sect.~\ref{sec:skasimulations}), we briefly discuss further emission properties by means of the modelled line profiles and line fluxes. As already mentioned, the \holine lines in the SKA1-Mid Band 5b are all similar with respect to spatial origin and velocities. Therefore, we show only the stacked profiles in Fig.~\ref{fig:profiles}. In this figure, we also show the result from two more models. The MHD\_B6 model is the same as  MHD\_B4 but with $\beta=10^6$. The MT\_HD\_LA model is the same as the MT\_HD model, but with an accretion luminosity of $L_\mathrm{accr}\approx10^{-2}\,\mathrm{L_\mathrm{\odot}}$, about two orders of magnitude lower than in the other models with $L_\mathrm{accr}\approx0.3\,\mathrm{L_\mathrm{\odot}}$. 

What is immediately apparent from Fig.~\ref{fig:profiles} is that the line profiles are very symmetric in velocity. The reason for this is that \holine emission is not absorbed by the dust, and hence we see the emission from almost everywhere in the disk where the line is excited. This is in contrast to other wind tracing lines, such as the \oiline where the red-shifted part of the line profile is often blocked by the dust. 

In the left panel of Fig.~\ref{fig:profiles}, which shows the stacked \holine lines for an inclination of zero degrees, differences in the full-width at half-maximum (FWHM) between models are clearly visible. As the disk is face-on, this broadening of the line is actually caused by the wind velocities. Hence, the models MHD\_B4 and MHD\_B6 show the highest velocities in the profile, but also the MT\_B4 model shows still significantly high-velocity wings, compared to the photo-evaporative models. Please note that those wings could be more pronounced in reality, as the inner disk radius for the MT models is as large as 0.5 au. The measured FWHM in the model profiles  range from about $\approx110\,\mathrm{km/s}$ for the MHD\_B4 model to $\approx25\,\mathrm{km/s}$ for the MT\_HD and PE models. In particular, one can see that the photo-evaporative models consistently show a narrower FWHM (by at least 10 km/s) compared to models that include magnetic fields. This is expected as pure PE models usually predict slower wind velocities. However, at higher inclinations (right panel, Fig.~\ref{fig:profiles}), almost all, but one, the MHD\_B4 model, show very similar profiles, with differences of about $\delta \mathrm{FWHM}\approx10\,\mathrm{km/s}$. Although the spectral resolution of SKA would allow us to discriminate between those models, identifying the dominant launching mechanism will be challenging. In particular, the PE model now shows a broader FWHM as, for example, the MT\_B4 and almost the same FWHM as the MHD\_B6 model. The reason for that is the Keplerian broadening of the lines, which becomes visible at higher inclinations. However, the general shape of the light profile might still provide information on the physical mechanism. For example, in the MHD\_B6 model, we see a shoulder at about $v=\pm20\,\mathrm{km/s}$ in the profile; such features can be spectrally resolved by SKA, if the line emission is strong enough.
\begin{figure}
    \centering
    \includegraphics[width=1\linewidth]{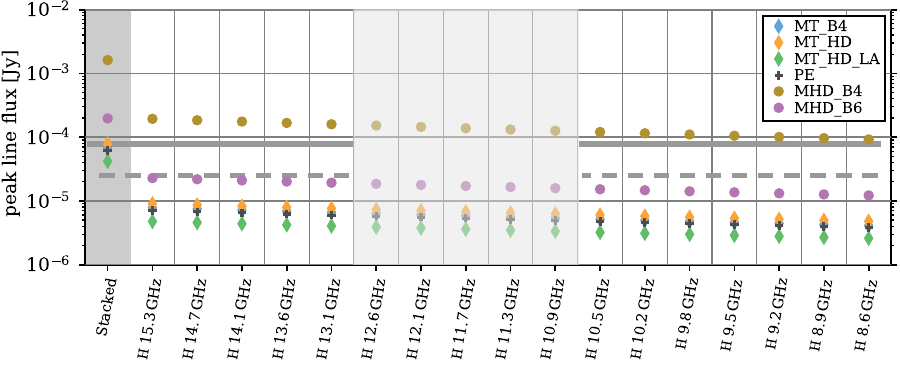}
    \caption{Predicted peak line fluxes for the \holine for all the here presented models at a distance of $140\mathrm{\,pc}$, inclination of $i=40^{\circ}$ and a bandwidth of $1\,\mathrm{km/s}$. On the left side of the panel, the stacked peak fluxes are shown (simply adding all the detectable lines). The horizontal grey solid and dashed lines indicate the detection limits, in terms of image rms, for integration times of 10 h and 100 h, respectively. The light-grey shaded area denotes the frequency range possibly impacted by (RFI), see Section \ref{sec:skasimulations}.}
    \label{fig:peakfluxes}
\end{figure}

Although the spectrally resolved line profiles of the \holine can provide new constraints for wind models, a detection of those lines with SKA-Mid is not guaranteed. In Fig.~\ref{fig:peakfluxes} we show the peak flux for all the modelled \holine lines relevant for Band~5b, with respect to sensitivity estimates computed using the SKA sensitivity calculator (see Section \ref{sec:skasimulations} for details on the computation). Only the MHD\_B4 model predicts high enough fluxes for the individual lines. However, stacking all those lines might also allow us to detect the weaker emission from the PE models. The much higher flux in the MHD\_B4 model is mainly a result of the high density in the wind. From the models we can derive an average Hydrogen number density within the main line emitting region (see Fig.~\ref{fig:magnetothermal} and Fig.~\ref{fig:MHDPE}). The MHD models show $n_\mathrm{H}\approx4-8\times10^8\,\mathrm{cm^3}$, whereas all the other models have about two orders of magnitude lower values close to $n_\mathrm{H}\approx5\times10^6\,\mathrm{cm^3}$. This huge difference is likely a consequence of the different modeling approaches; however, generally speaking, it is expected that \holine would trace lower-density but possibly hotter (less efficient cooling) gas in thermally driven winds, whereas only magnetically driven winds can produce high-density outflows at the inner edge of the disk.

\begin{figure}
    \centering
    \includegraphics[width=\linewidth]{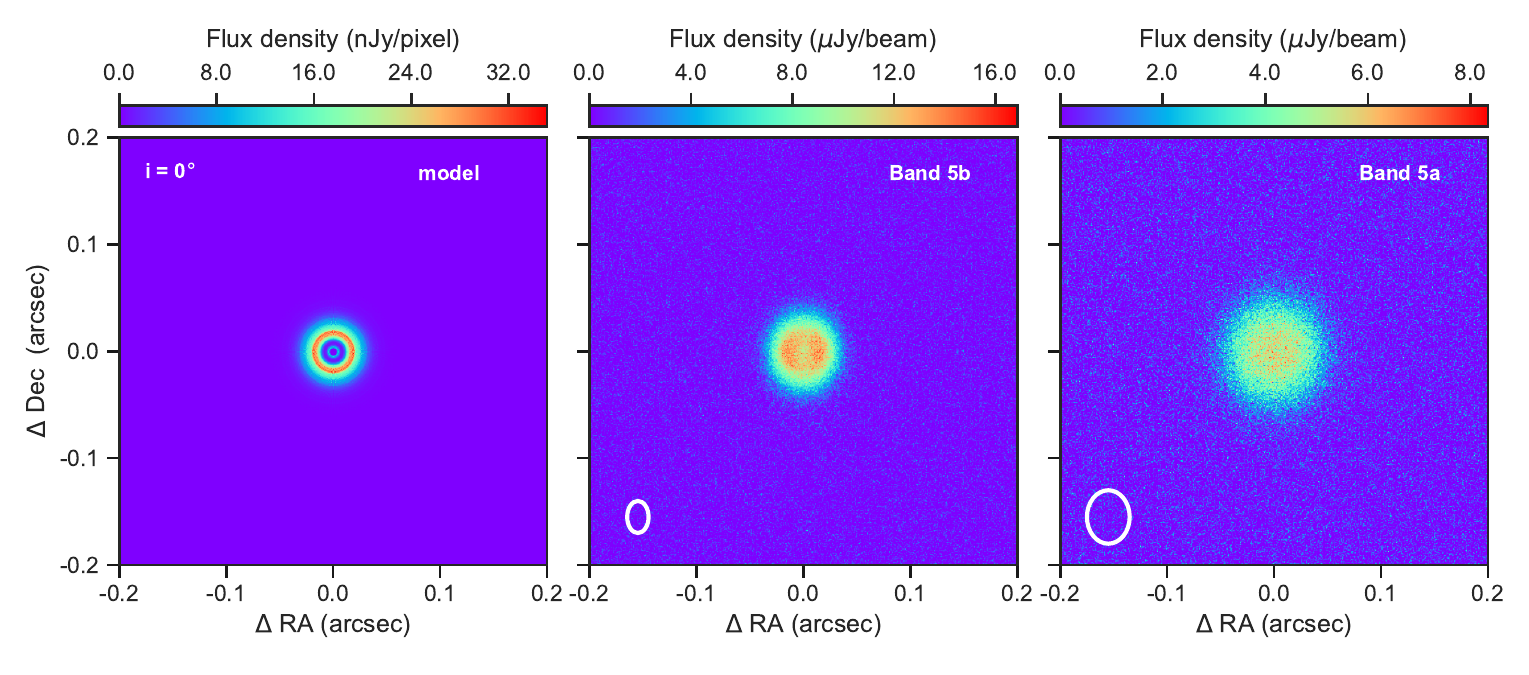}
    \includegraphics[width=\linewidth]{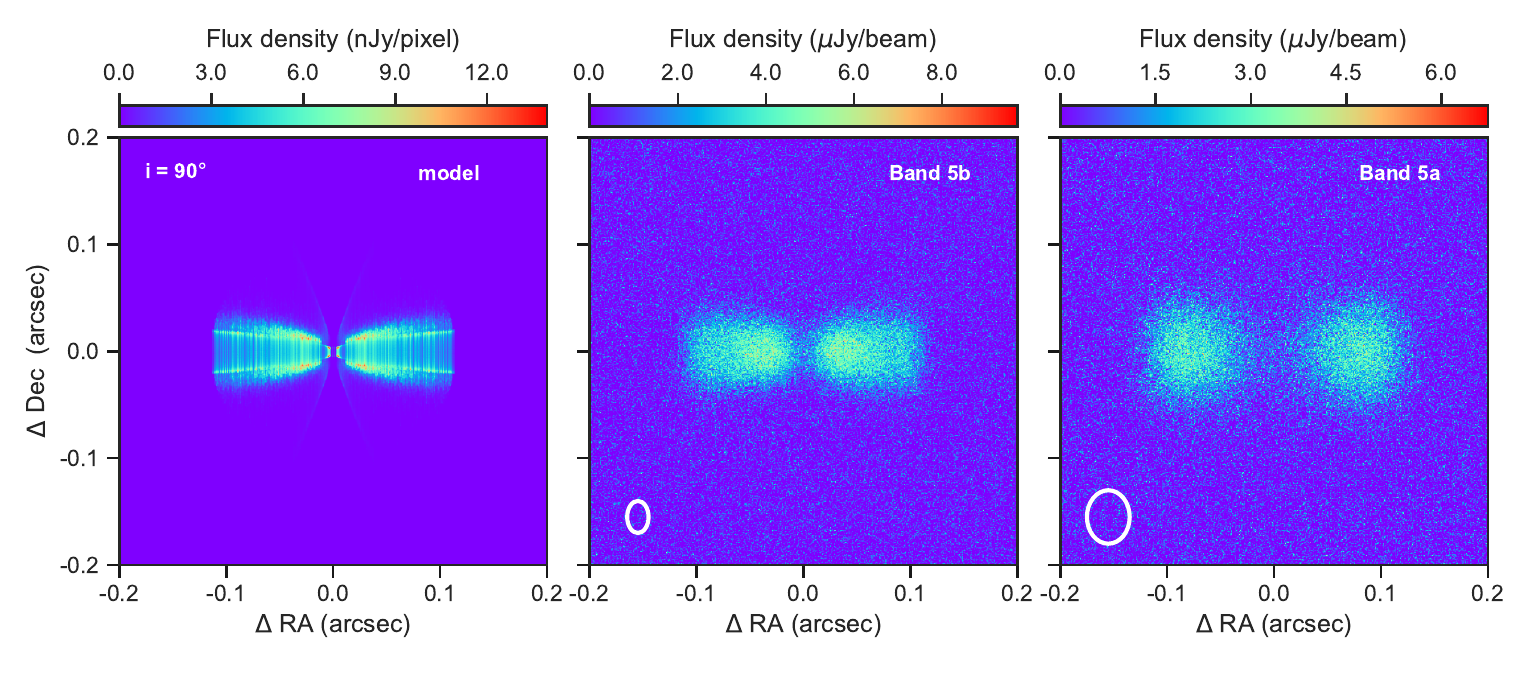}
    \caption{Left column: simulation of free--free emission with the MT\_B4 model at 2.5\,cm, for a disk with an inclination of 0$\degree$ (top) 
    and 90$\degree$ (bottom). Middle and right panels: simulated observations for 10h integration time with SKA-Mid AA4 at Band~5a and 5b, 
    using briggs weighting with robust = -2. The resulting beams are 0\farcs03 $\times$ 0\farcs02 and 0\farcs05 $\times$ 0\farcs04, respectively.}
    \label{fig:skasim-ff-mtb4}
\end{figure}

\subsection{SKA simulated observations}
\label{sec:skasimulations}

\begin{figure}
    \centering
    \includegraphics[width=0.49\linewidth]{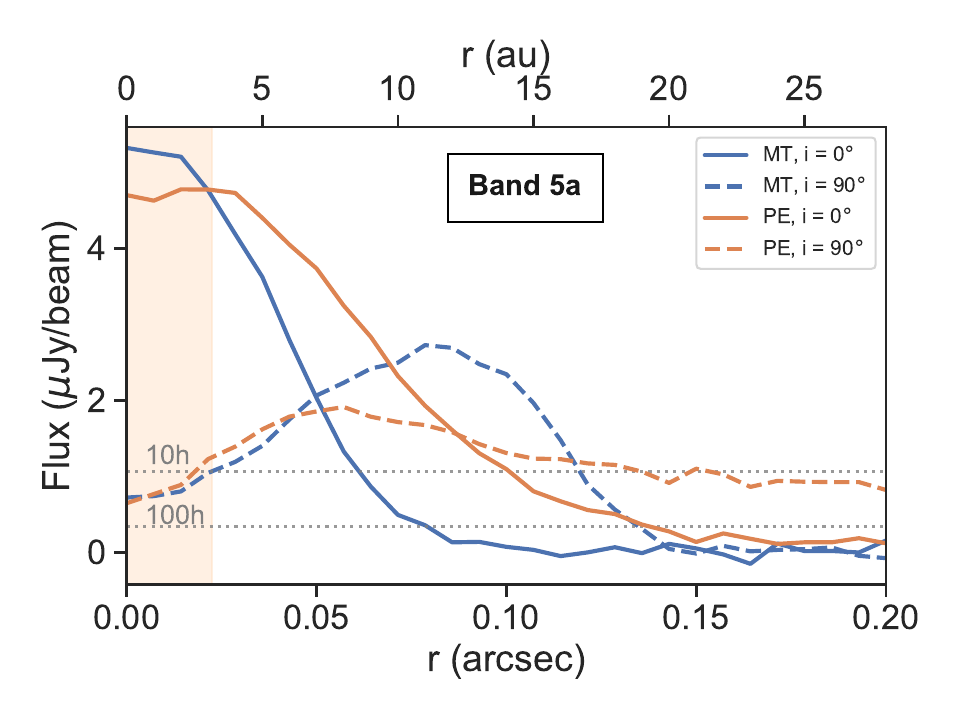}
    \includegraphics[width=0.49\linewidth]{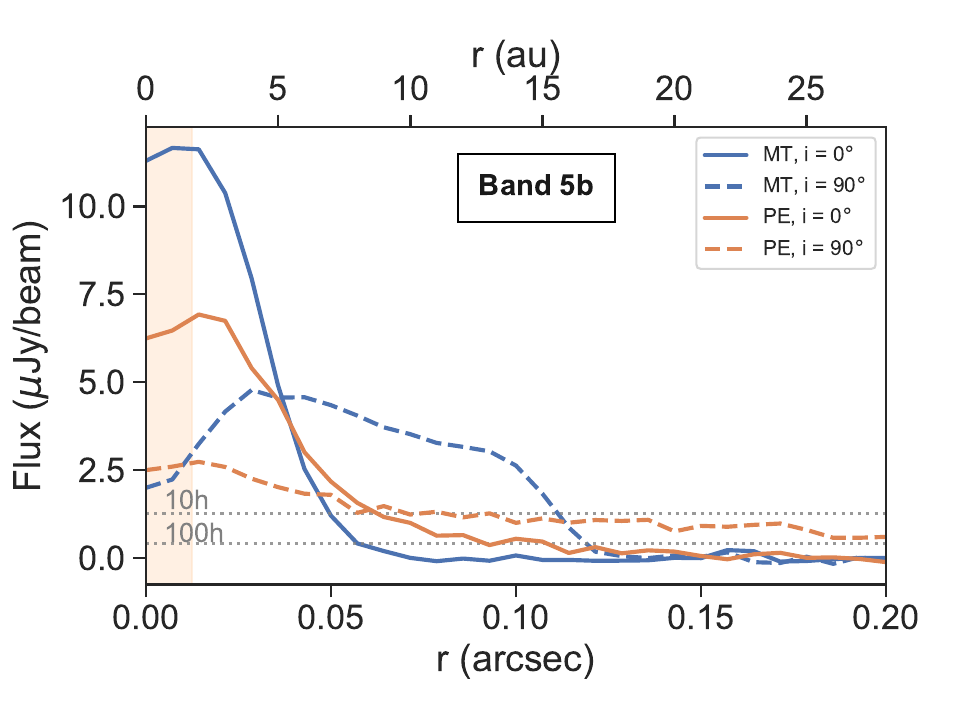}
    \caption{Radial profile along the horizontal direction of the free--free emission from simulations of SKA observations with 10 hours integration time and briggs weighting with a robust of -2, from the MT\_B4 (see Figure \ref{fig:skasim-ff-mtb4}) and PE models at 4.5\,cm (Band 5a, left panel) and 2.5\,cm (Band 5b, right panel), for disk inclinations of 0$\degree$ and 90$\degree$. Dotted horizontal lines show the rms value for 10 hours and 100 h integration time, the shaded area correspond to half the beam average FWHM.}
    \label{fig:skasim-ff-prof}
\end{figure}

To simulate SKA-Mid AA4 observations, we used the SKA Sensitivity Calculator on a general target at a RA of 15$h$00$m$00$s$ and Dec -25$d$00$m$00$s$, corresponding approximately to a source in the Upper-Sco Star Forming Region. We computed the sensitivity in continuum mode (3.9~GHz and 5~GHz bandwidth for Band 5a and Band 5b, respectively) in the AA4 configuration with 133 antennas with a 15\,m diameter, to obtain the predicted noise of the final image and the synthesized beam. The final image was produced by adding the noise (as a gaussian random distribution with a $\sigma$ corresponding to the rms) to the model sky brightness convolved with the final beam. 
In Figure \ref{fig:skasim-ff-mtb4} we show the model results and SKA simulated observations of the free--free emission of a solar-mass star at 140\,pc, at inclinations of 0$\degree$ and 90$\degree$ (see previous Section~\ref{sec:modeldescription}). In this example we show how we can detect and resolve the emission at SKA-Mid Band 5a and 5b, with 10 hours integration time, reaching a rms of about 1.1 and 1.3 $\mu$Jy/beam and achieving a peak SNR of about 4--8 and 8--13 (depending from the disk inclination), respectively. The intensity profiles for a radial cut are shown in Figure \ref{fig:skasim-ff-prof}. Although the emission from the MT models can reach a higher peak in the resolved profiles, it is more compact than the PE emission and has a lower integrated flux density at all cm-frequencies analysed here, as discussed in the next paragraph.

\begin{figure}
    \centering
    \includegraphics[width=\linewidth]{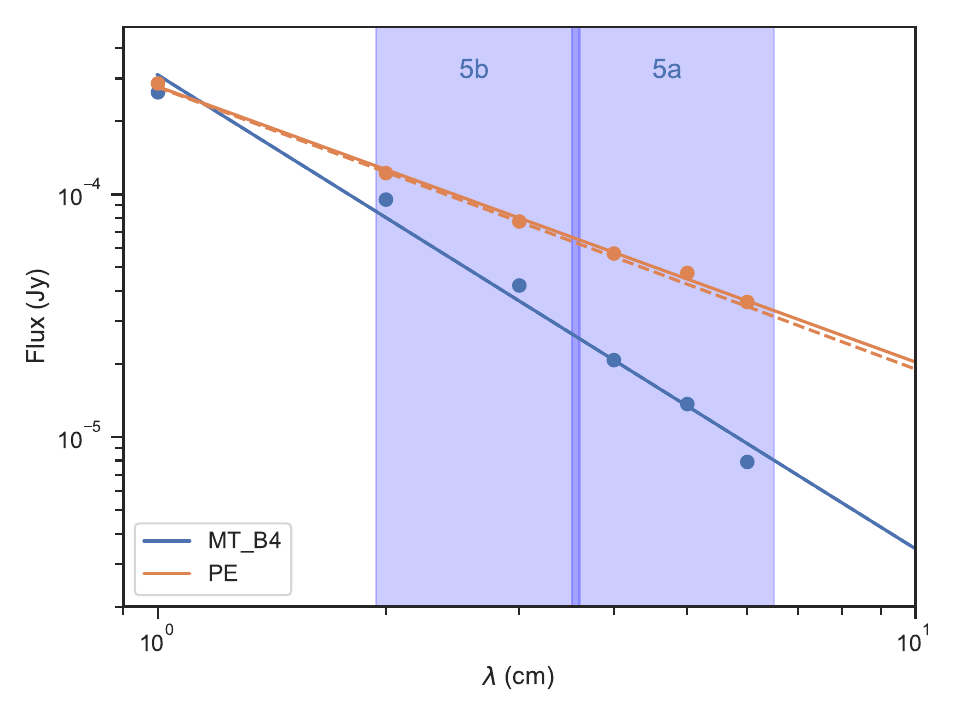}
    \caption{Spectral Energy Distribution of the free--free emission from the magneto-thermal model (orange circles) and the photoevaporative model (blue circle). The solid lines show the interpolated power-laws between the flux and the frequency (similarly for the dashed lines, computed using the models with inclination of 90\degree). Shaded blue areas show the frequencies covered by SKA-Mid band~5. }
    \label{fig:sed}
\end{figure}

Figure \ref{fig:sed} shows the spectral energy distribution of the free--free emission corresponding to the photoevaporation-only and magneto-thermal models. While the integrated fluxes predicted at 1\,cm are very similar between the two models, they progressively diverge going to longer wavelength, reaching a factor of $\sim$4 in SKA Band 5a. Furthermore, the magneto-thermal model displays a steeper slope with a spectral index of 1.97, compared to a value of 1.13 for the PE model, values that indicate a fully and partially optically thick emission, respectively. 
We note here that although free--free emission from Class~II disks is often assumed to be optically thin at the frequencies considered here, the 4--400\,GHz survey of 32 Class~II disks in the Taurus-Aurigae region conducted by \citet{Chung2025} revealed that most sources exhibit an optically thick free--free emission at frequencies $>$6\,GHz.

Detecting and spectrally resolving Hydrogen recombination lines from disks could provide a powerful diagnostic of the mechanism giving rise to ionized gas emission. 
As shown in Section \ref{sec:modeldescription}, simulations of recombination lines raising from MHD winds predict a significantly higher flux, with respect to photoevaporation-only winds (although this is connected to the higher gas density in the former). 
Simulating 100 hours integration time with SKA 
indicates that we would be able to detect only the set of H$\alpha$ lines at SKA-Mid band~5 predicted by the MHD\_B4 model. A spectral resolution of 0.25~km/s would lead a peak SNR of 6, while spectral binning to a velocity resolution of 1km/s would give a peak SNR of 10 and still allow to resolve the line profile. 

\begin{figure}
    \centering
    \includegraphics[width = 0.7\linewidth]{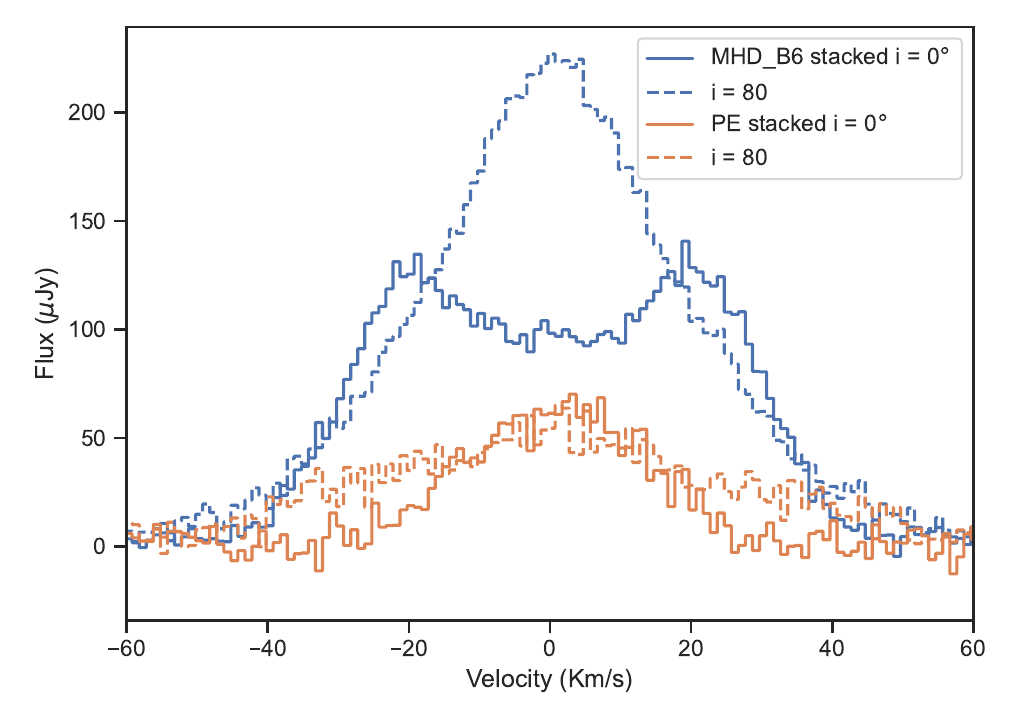}
    \caption{Spatially integrated spectra of the 12 stacked H$\alpha$ lines from the PE model and the MHD\_B6 model, for disk inclinations of 0$\degree$ and 80$\degree$. The cubes have been obtained by assuming an integration time of 10 hours and using robust 2 parameter with 0\farcs024 tapering and a spectral resolution of 1km/s.}
    \label{fig:skasim-pe-mhd_b6-pe}
\end{figure}

A promising strategy comes from stacking the hydrogen recombination lines within the SKA-Mid Band 5 range, which is justified by our simulations indicating that these lines originate from the same region within the system. 
We simulated this by adding up the 12 H$\alpha$ lines in Band 5b, where we excluded the lines falling in the range between 10.7 and 12.7 GHz, because of RFI\footnote{\href{https://www.skao.int/en/news/198/skao-needs-corrective-measures-satellite-mega-constellation-operators-minimise-impact-its}{SKAO needs corrective measures from satellite `mega-constellation`.}} (Radio Frequency Interferences). We considered an integration time of 10 hours with a spectral resolution of $0.25\,\mathrm{km/s}$ and a robust parameter of 2 for all the five models presented in Section \ref{sec:modeldescription}, obtaining stacked cubes where the final noise is reduced by a factor of $\sqrt{n}$ where $n$ = 12, and corresponds to 59 $\mu$Jy/beam with a beam of 0\farcs1. 
For all models we obtain a detection with a peak signal-to-noise ratio of at least 6, in particular for PE, MT\_B4, MT\_HD, MHD\_B6 and MHD\_B4 we get peak SNR values of 6, 10, 11, 12 and 31, respectively. A spectral binning to a resolution of 1~km/s can further decrease the noise by a factor 2 and increase the peak SNR by the same amount, while still resolving the line profiles in all models. In Figure \ref{fig:skasim-pe-mhd_b6-pe} we show an example of how the stacked cube of H$\alpha$ lines would appear for 10 hours integration and a spectral resolution of 1\,km/s for the pure photoevaporation model (PE) and the MHD\_B6 model, for two different disk inclinations. The simulated observations would allow the different line profiles to be distinguished, with the MHD models showing higher line intensities as discussed in section \ref{sec:modelresults}.

The SKA-Mid array is planned to 
operate in combination with the 64  MeerKAT\footnote{https://www.sarao.ac.za/science/meerkat/} antennas, each 13.5\,m in diameter, a precursor to the SKA telescope. 
Integrating this array into SKA-Mid AA4 would enhance the final sensitivity of narrow band observations by a factor of about 1.4. The gain in continuum sensitivity will be slightly smaller, since MerKAT band 5b is expected to provide a 2.5~GHz bandwidth, as opposed to the 5~GHz bandwidth of SKA-Mid Band 5b, yielding an estimated gain of about a factor 1.2 in sensitivity. 
These are approximate estimates that are done assuming  MeerKAT system temperatures comparable to those of SKA-Mid, and that the full 5\,GHz bandwidth is correlated for the SKA-Mid baselines. 

\section{Prospects for SKA surveys}
The large FOV of SKA-Mid antennas, corresponding to about 6.7 arcmin at Band 5b and 12.5~arcmin at Band 5a, will enable simultaneous observations of multiple targets in nearby star forming regions. 
When observing Class II disks, the spatial scales and sky brightness are modest, and we showed that the typical integration times for SKA can be of the order of tens of hours to observe the ionized gas emission. 
Broad band observations with the SKA of continuum emission towards young disks would provide valuable information for a variety of science cases. In the context of this chapter, detection of free-free emission could be obtained with a few hours integration time: we showed in Section \ref{sec:skasimulations} that in one hour we can detect free-free with a SNR of 5 observing a system at 140\,pc, while with 10~hours integration time we can spatially resolve the free--free emission predicted by our simulations.
While in Band 5b we could probe 
the full disk population of
a star forming region like Ophiuchus in 60 pointings \citep[see the chapter by][]{Garufi01.2026.SKA}, 
in Band 5a the number of pointings would reduce to about 20. Therefore, a survey aimed at detecting free--free emission from a star forming region at 140\,pc would take approximately 60 and 20~hours of SKA time in Band 5b and 5a, respectively. 
Such observations would constrain both the free--free and the dust component, as discussed in the chapters by \citet{Garufi01.2026.SKA} and \citet{Wu01.2026.SKA}. 

As shown in Section \ref{sec:skasimulations}, spectral information is essential for studying the mechanisms underlying radio emission in disks: different winds models (photoevaporation-only and magnetised) predict distinct spectral slopes in the cm-range and in particular across SKA-Mid band 5a and 5b. Hence, simultaneous multi-band observations at these frequencies would be essential to disentangle the origin of continuum emission. 

A similar argument applies to high spectral resolution windows: we demonstrated in Section \ref{sec:skasimulations} how stacking the Hydrogen recombination lines in Band 5b would allow us to obtain a spectrally resolved detection for all simulated models. An observing mode that allows simultaneous coverage of several high resolution spectral windows across Band 5 would offer a unique opportunity to detect for the first time Hydrogen recombination lines from a disk wind in a Class II disk. For the models presented here, the spectral resolution results essential to distinguish between the different wind launching mechanisms, and indirectly infer disk physical conditions, such as gas density and temperature, through direct comparison of the observed line profiles with the theoretical predictions. 

As described in Section \ref{sec:radioemission}, contamination from non-thermal processes can be significant at these wavelengths \citep[e.g.][]{Coutens2019}.  Therefore, it is essential to perform multi-epoch observations to track the variability of the non-thermal emission and to identify the ``quiet epochs'' with low non-thermal contribution, when the dust emission and the free-free emission can be easily disentangled. 

\section{Synergies with other facilities}
\label{sec:synergies}
\subsection{Synergies with radio facilities at millimeter-centimeter wavelengths}

As mentioned in Section \ref{sec:models}, hydrogen radio recombination lines are expected to present larger peak intensities with increasing frequency since their flux varies as $S_\nu\propto\nu^{1.1}$. RRLs have indeed been observed at millimeter wavelengths in massive star-forming regions with known UC HII regions presenting very broad line profiles indicative of ionized winds \citep{Jaffe1999}. Some of these objects even present non-thermal amplification of the RRL lines in masers \citep{MartinPintado1989,JimenezSerra2011,JimenezSerra2013,MartinezHenares2024}. The bright RRL masers are excellent tools to probe the innermost regions in these massive objects. By comparing the accurate 2D centroid maps obtained from the RRL maser lines with the 3D radiative transfer code MORELI \citep{BaezRubio2014}, RRL masers provide exquisite information about the kinematics and physical structure of the ionized gas distributed in disks, winds and jets \citep{MartinPintado2011}. This method has been successfully used recently for the MWC349A massive line star and has revealed the presence of a high-velocity jet besides the already known ionized disk and disk wind \citep{MartinezHenares2023,MartinezHenares2024}. 

In contrast to their massive counterparts, no millimeter RRL has been reported toward low-mass protostars to date. Therefore, the high-sensitivity and imaging fidelity of the SKA will allow the detailed study of the kinematics and physical structure of disk winds and jets by using centimeter RRLs. Synergies with ALMA WSU are foreseen since the ALMA bandwidth upgrade will allow the simultaneous observations of multiple millimeter RRLs enabling stacking of these lines also at millimeter/submillimeter wavelengths, enhancing the probability of detection. While millimeter RRLs probe the inner and higher-density regions in ionized disks,  winds and jets, centimeter RRLs trace the lower-density regime of the ionized gas. Hence, the combination of low-$n$ and high-$n$ quantum number transitions of RRLs will provide the complete 3D structure of winds and jets in both low-mass and high-mass protostars \citep[see also the chapter led by][]{Sabatini01.2026.SKA}.

\subsection{Synergies with GRAVITY+}

The Square Kilometre Array will provide, for the first time, a systematic view of ionised gas 
emission in disks at centimeter wavelengths, tracing free--free emission and hydrogen recombination 
lines on scales of tens to hundreds of au. Yet, many of the most critical processes shaping disk 
evolution occur within a few astronomical units of the star, where accretion flows are established 
and winds and jets are launched. It is precisely this innermost region that will be accessible with 
the upcoming GRAVITY+ interferometric facility at the VLTI. 

GRAVITY+ delivers milli-arcsecond angular resolution and high spectral sensitivity in the 
near-infrared, directly resolving the launching regions of ionised winds and accretion flows 
traced by Br$\gamma$, He~I, and other hydrogen recombination lines. These capabilities are highly 
complementary to SKA, which will simultaneously probe the extended ionised structures through 
cm-wavelength free--free emission, and recombination line diagnostics. Together, the two facilities 
bridge spatial scales from sub-au to hundreds of au, enabling a unified picture of accretion and 
ejection across the entire disk.

\textit{Sensitivity is central to this synergy.} The new GPAO+LGS adaptive optics upgrade of 
GRAVITY+ extends the accessible sample of massive young stellar objects by nearly an order of 
magnitude, pushing the K-band sensitivity limit for fringe tracking to $\sim$12.5\,mag on-axis and 
opening access to deeply embedded sources. This leap means that hundreds of MYSOs, previously 
out of reach for near-IR interferometry, can now be observed at milli-arcsecond resolution. At the same 
time, SKA will reach sub-$\mu$Jy sensitivities in the cm band, enabling detection of faint free--free 
emission and hydrogen recombination lines from photoevaporative or magnetised winds that are beyond the reach of current VLA or ATCA surveys. The combination of these sensitivity gains is 
transformative: SKA can detect extended ionised winds at large scales, while GRAVITY+ resolves 
the innermost launching regions, together allowing us to track the same processes continuously 
across spatial scales and evolutionary stages. 

This synergy is particularly compelling in the context of massive young stellar objects (MYSOs). 
Recent interferometric studies with GRAVITY and AMBER have revealed dynamical inner disks, 
spiral substructures, and variable emission lines in massive accretion disks \citep[e.g.][]{Kraus2010, 
Caratti2016, Frost2021, Koumpia2021}. Numerical simulations predict that such inner disks may be 
gravitationally unstable, fragmenting into clumps, and producing highly time-variable accretion 
bursts \citep[e.g.][]{Elbakyan2023}. These events alter the geometry and emission 
properties of the inner disk on timescales of months to years. SKA observations of recombination 
lines and continuum variability, combined with time-resolved interferometric information from 
GRAVITY+, will make it possible to directly link changes in the extended ionised wind to 
instabilities and bursts occurring at the disk surface.

Synergies are also central for binarity. GRAVITY+ will, for the first time, deliver systematic 
statistics of MYSO binaries down to separations of a few au, thanks to the new GPAO+LGS upgrade 
that expands the accessible target sample by an order of magnitude. At the same time, SKA will 
resolve circumbinary ionised structures and quantify how companions perturb the launching and 
collimation of outflows. Joint observations will therefore place binary formation and multiplicity 
in the broader context of accretion and feedback.

In summary, SKA and GRAVITY+ together provide a uniquely powerful, multi-scale perspective on 
disk evolution. GRAVITY+ reveals the innermost au-scale physics of accretion and wind launching, 
while SKA maps the extended ionised environment and its variability. This combination will make 
it possible to connect small-scale instabilities and bursts to large-scale dispersal, to link 
magnetic field structures across many orders of magnitude in scale, and to anchor binarity studies 
within the broader context of disk and wind evolution. The integration of these facilities will 
open a new chapter in our understanding of how disks evolve, disperse, and give rise to stars and 
planetary systems. 

\subsection{Synergies with JWST}
\label{sec:synJWST}
Observations made with the James Webb Space Telescope (JWST) enable high-quality studies of ionized winds at infrared wavelengths, thanks to its extraordinary sensitivity and the multi-range spectroscopy offered by the NIRSpec and MIRI-MRS instruments on board. The JWST is capable of providing unprecedented spectro-imaging data in key diagnostic emission lines in both the near-infrared and mid-infrared ranges: hydrogen recombination lines and forbidden lines of interest such as [Ne II], [S III], [Ar II], [AR III] provide spatially resolved detections of partially ionized winds \citep[e.g.][]{Bajaj2024,Pascucci2025}; transitions of molecules such as H$_\mathrm{2}$ and CO provide information about the inner regions of the disc and the cavities where ionised winds flow \citep{Arulanantham2024,Francis2024,Franceschi2024}. 
The JWST's ability to spatially and spectrally resolve these signatures allows for the study of line profiles, intensity ratios, and relative velocities. In this way, JWST can provide information on the radiation environment at the disk surface and on the spatial distribution, chemical composition, temperature, and ionization state of the warm gas component of the winds. In turn, these investigations can help assess whether the winds are driven primarily by EUV photoevaporation, X-rays, FUV photoevaporation, or by magnetothermal processes \citep{Espaillat2023, Sellek2024,Schwarz2025}. 
In this context, the SKAO will provide a new and fundamental window for the study of ionized gas, complementary to the JWST, as these facilities probe different physical components of the same winds. JWST observations focus on the hot, partially ionized layers of gas at the surface of the disk and within shocks along the outflow, while the SKA traces the fully ionized component of the outflow on different scales. 
The free--free continuum emission and radio recombination lines studied with SKA are directly sensitive to the density of the ionized gas and therefore provide reliable measurements of mass loss rates and wind geometry. Thus, while JWST characterizes the origin and excitation of the wind, SKA measures its total ionized mass and global energy, allowing us to determine how efficiently the winds disperse disk material. An important complementary contribution from SKA will come from its higher spectral resolution compared to JWST. Indeed, JWST's resolution limit of 50-80 km s$^\mathrm{-1}$ limits our ability to identify important properties of the flow, such as the transverse velocity structure, which can reveal a layered kinematic structure of the wind and its rotation around the axis. These properties are fundamental for testing the nature of the acceleration mechanism and the connection of the winds with the disc structure, as illustrated by recent studies conducted with ALMA on molecular flows \citep{Lopez-Vazquez2024, Bacciotti2025}. 
Integrating the sensitive JWST observations of ionized winds with the accurate kinematic analysis provided by SKA on recombination lines will allow an unprecedented examination of the applicability of models for the origin of winds. It should also be noted that the sample size for disks with JWST with robust detections of ionized gas remains small (a few objects), so it is not yet clear how universal such winds are among different types of disks, masses, ages, and inclinations. In this regard, the SKA's ability to detect traces of ionized winds could multiply the number of systems to be examined with both facilities, providing a unique opportunity to build a robust statistic of the observed phenomena.

\section{Summary}
The ionized gas component of protoplanetary disks provides key insights on the mechanisms driving disk evolution. Despite the recent advancements in observing capabilities, detecting and characterizing such ionized component remains an observational challenge. 

We have performed state-of-the-art simulations of photoevaporative, magneto-thermal, and MHD winds, to generate predictions for SKA-Mid future observations and detection potential of wind tracers. 
Our results indicate that SKA-Mid will be capable of resolving free--free emission in nearby disks and possibly detect hydrogen recombination lines, allowing us to probe wind kinematics and ionization conditions. 
The comparison between the photoevaporative and magneto-thermal model for the free--free emission indicates a lower flux and steeper spectral slope for the latter at cm wavelengths, with a more pronounced contrast  in Band~5a of SKA-Mid. These flux levels would be accessible with only a few hours of integration with SKA-Mid.
Regarding Hydrogen RRLs, the differences among the analysed models suggests that line widths and spatial morphologies are discriminants between thermally- and magnetically-driven winds. While SKA-Mid is unlikely to spatially resolve such lines,  stacking Band~5b transitions would enable to detect and spectrally resolve them with approximately 10 hours integration time. 

Synergies with facilities operating at complementary wavelengths, such as ALMA, ngVLA, JWST, and GRAVITY+, will be key to link the inner wind-launching regions to the larger scale outflows and the properties of the solid material, providing a comprehensive and multi-scale view of protoplanetary disks and shed light on the processes driving disk evolution and setting the stage for planet formation.  

\section*{Acknowledgments}
{\footnotesize{\it This project has received funding from the European Research Council (ERC) under the European Union's Horizon Europe research and innovation program (grant agreement No. 101053020, project Dust2Planets). I.J-.S acknowledges funding from grant PID2022-136814NB-I00 funded by the Spanish Ministry of Science, Innovation and Universities/State Agency of Research MICIU/AEI/10.13039/501100011033 and by ``ERDF/EU''. JDI acknowledges support from an STFC Ernest Rutherford Fellowship (ST/W004119/1). Y.W. acknowledges the EACOA Fellowship awarded by the East Asia Core Observatories Association. This research was supported by the funding from the National SKA Program of China under Grant No. 2025SKA0120100. E.B. acknowledges financial support from the Italian Ministry of University and Research (MUR) under the Italian Science Fund (FIS 2 Call -- Ministerial Decree No. 1236 of 1 August 2023). C.C., L.P. and G.S. acknowledge financial support under the National Recovery and Resilience Plan (NRRP), Mission 4, Component 2, Investment 1.1, Call for tender No. 104 published on 2.2.2022 by the Italian Ministry of University and Research (MUR), funded by the European Union - NextGenerationEU - Project Title 2022JC2Y93 Chemical Origins: linking the fossil composition of the Solar System with the chemistry of protoplanetary disks - CUP J53D23001600006 - Grant Assignment Decree No. 962 adopted on 30.06.2023 by the Italian Ministry of Ministry of University and Research (MUR); the project ASI-Astrobiologia 2023 MIGLIORA (``Modeling Chemical Complexity'', F83C23000800005); the INAF-GO 2024 fundings ICES, the INAF-GO 2023 fundings PROTOSKA (``Exploiting ALMA data to study planet forming disks: preparing the advent of SKA'', C13C23000770005); the INAF Mini-Grant 2022 ``Chemical Origins'' (PI: L. Podio) and the INAF Minigrant 2023 TRIESTE (``TRacing the chemIcal hEritage of our originS: from proTostars to planEts''; PI: G. Sabatini). CHR, BE and MW acknowledge the support of the Deutsche Forschungsgemeinschaft (DFG, German Research Foundation) Research Unit ``Transition discs'' - 325594231. This research was supported by the Excellence Cluster ORIGINS, which is funded by the Deutsche Forschungsgemeinschaft (DFG, German Research Foundation) under Germany's Excellence Strategy - EXC-2094 - 390783311. CHR is grateful for support from the Max Planck Society. }}











\bibliographystyle{abbrvnat-maxbibnames4}
\bibliography{chapter} 

\end{document}